\documentclass[twocolumn,aps, prd, amsmath, amssymb, bm]{revtex4-1}
\usepackage{times}
\usepackage{graphicx}
\usepackage{verbatim}
\usepackage{tikz}
\usepackage{color}

\usepackage[T1]{fontenc}
\usepackage{aecompl}
\usepackage{calligra}
\DeclareMathAlphabet{\mathcalligra}{T1}{calligra}{m}{n}
\DeclareFontShape{T1}{calligra}{m}{n}{<->s*[2.2]callig15}{}

\usepackage{hyperref}
\hypersetup{
    pdfnewwindow=true,      
    colorlinks=true,       
    linkcolor=blue,          
    citecolor=blue,        
    filecolor=blue,      
    urlcolor=blue           
}

\def\apj{ApJ}                 
\def\apjl{ApJL}               
\def\apjs{ApJS}               
\def\mnras{MNRAS}             
\def\aap{A\&A}                
\def\aj{AJ}                   
\def\nat{Nature}              
\def\nar{NewAR} 			  

\def\bin{\rm{bin}}

\def\orb{\rm{orb}}

\def\res{\rm{res}}

\def\min{\rm{min}}
\def\max{\rm{max}}
\def\Edd{\rm{Edd}}

\def\Msun{{M_{\odot}}}

\def\be{\begin{equation}}
\def\ee{\end{equation}}

\def\bea{\begin{eqnarray}}
\def\eea{\end{eqnarray}}

\def\SBHB{{\rm SBHB}}

\begin{document}
\title[]{Detecting the Orbital Motion of Nearby Supermassive Black Hole Binaries with \textit{Gaia}}
\author{Daniel J. D'Orazio}
\email{daniel.dorazio@cfa.harvard.edu}
\affiliation{Department of Astronomy, Harvard University, 60 Garden Street Cambridge, MA 01238, USA}

\author{Abraham Loeb}
\affiliation{Department of Astronomy, Harvard University, 60 Garden Street Cambridge, MA 01238, USA}

\begin{abstract} 

We show that a 10 year \textit{Gaia} mission could astrometrically detect
the orbital motion of $\sim 1$ sub-parsec separation supermassive black hole
binary in the heart of nearby, bright active
galactic nuclei (AGN). Candidate AGN lie out to a redshift of $z=0.02$ and in the
V-band magnitude range $10 \lesssim m_V \lesssim 13$. The distribution of
detectable binary masses peaks at a few times $\sim 10^7 \Msun$ and is truncated above a few times $\sim 10^8 \Msun$.

\end{abstract}

\maketitle

\section{Introduction}

The \textit{Gaia} satellite is mapping the positions of the stars with
unprecedented precision. Its 5 year mission: to survey the 6D phase space
coordinates of a billion stars to an astrometric precision of a few $\mu$as
\citep{GaiaI:2001, GaiaII:2016, GaiaDR2:2018}. \textit{Gaia} will observe not
only stars, but all optical sources brighter than an apparent magnitude of
$\sim20$. This includes active galactic nuclei (AGN), namely distant and
powerful sources of multi-wavelength emission driven by gas accretion onto
supermassive black holes (SBHs) at the centers of galaxies.

AGN are used to calibrate \textit{Gaia} astrometric position measurements,
both via \textit{Gaia}'s optical astrometry as well as with radio-frequency
VLBI \citep{GaiaDR2:astrmsoln:2018}. The AGN are chosen as calibrators because
they are distant and hence expected to exhibit very little proper motion or
parallax. Despite this expectation, \textit{Gaia} has detected $\gtrsim 1$mas
offsets in optical and radio positions of AGN, probing dislodged AGN or
radio/optical jet properties \citep{Makarov+2017, PetrovKovalev:2017,
KovalevPetrov:2017, PetrovKovalev:2018}. In this \textit{Letter} we show that
on $\lesssim 50\mu$as scales, this expectation is also relevant for AGN that
harbor sub-parsec (pc) separation SBH binaries (SBHBs). Orbital motion of one
or both accreting SBHs in a SBHB can change the position of the optical
emitting region of the AGN by an angle greater than the astrometric precision
of \textit{Gaia}. SBHB orbital motion would
be distinct from the linear motion expected for a jet or ejected AGN. Because
binary-induced motions will only occur for a minority of AGN, there will be
little impact on \textit{Gaia}'s calibration. This observation does, however,
present a path towards definitive detections of sub-pc separation SBHBs.

While solid lines of evidence lead us to expect that SBHBs reside in the
centers of some galaxies \citep{Begel:Blan:Rees:1980}, their definitive
detection at sub-pc separations is yet to be obtained. The existence of sub-pc
SBHBs is of special importance as it embodies the `final-parsec
problem' \citep{Begel:Blan:Rees:1980, MerrittMilos:2005:LRR}, determining the
fate of SBHBs. If interaction with the environments in galactic nuclei can
drive SBHBs to sub-pc separations, then they will merge via emission of
gravitational waves (GWs), detectable out to redshifts $z \geq 10$ by the
future space-based GW observatory LISA \citep{LISA:2017}, and generating a
low-frequency stochastic GW background detectable by the Pulsar Timing Arrays
\citep[PTAs;][]{PTAs}.

To determine which, if any, proposed mechanisms,
\cite[\textit{e.g.},][]{GouldRix:2000, ArmNat:2002:ApJL, MacFadyen:2008,
Goicovic+2016, Gualandris+2017:TriaxRefill, Yike+2017}, solve the final-parsec
problem in nature, one must characterize a population of sub-pc SBHBs. Current
detection methods are indirect and require campaigns that last many years
\citep[\textit{e.g.},][]{ShenLoeb:2010, Tsalmantza:2011, Bogdanovic+2009,
Eracleous+2012, McKFeZoltan:2013, DecarliDott:2013:SpecMBHBcandI, Shen+2013I,
Liu+2014II, LiuEracHalp:2016, NguyenBogdan+2018, HKM09,
DHM:2013:MNRAS, PG1302MNRAS:2015a, Farris:2014, Graham+2015b, Charisi+2016,
LiuGezariPLCs+2016, DZ:2017, DDLens:2017, Gower+1982, Roos:1993,
MerrittEker:2002, Zier:2002, Romero:2000, Kun+2014, Kun+2015:PG1302,
KulkLoeb:2016, LiuChen:2009, StoneLoeb:2011, Coughlin+2017}. While these
techniques provide a way towards identifying and vetting SBHB
candidates via a combination of indirect methods, a more direct approach is
desired.

Recently, we have shown that mm-wavelength VLBI possesses the astrometric
resolution and longevity to repeatedly image SBHB orbits out
to redshift $z\sim0.5$, providing direct evidence for SBHBs in radio-loud AGN
\citep{D'OrazioLoebVLBI:2018}. The technique that we propose here also
directly tracks the SBHB orbit with the advantage that target AGN need
not be bright in mm-wavelengths and that unlike VLBI, \textit{Gaia}
is conducting a survey mission that will map the entire sky, and, as
we show, could find evidence for SBHBs within the next $5-10$ years.

\begin{table*}
\begin{tabular}{l|l|l|l|l}
  Parameter         & Meaning                              & Fiducial & Optimistic & Pessimistic  \\
\hline
\hline
  $f_{\bin}$       & The fraction of AGN harboring SBHBs & $0.1$ & " & " \\
  $f_{\Edd}$       & The Eddington fraction of bright AGN & $0.1$ & " & " \\
  $BC$             & Bolometric correction from V-band & $10.0$ & " & " \\
  $t_{Q}$          & The AGN lifetime & $10^7$~yrs & $5 \times 10^6$~yrs & $10^8$~yrs \\
  $V-I_c$          & A mean color for nearby AGN & $0.7$ & $1.1$ & $0.0$ \\
  $P_{\max}$       & Mission lifetime  & $10$~yrs \ ($5$~yrs)  \ ($20$~yrs) & $10$~yrs \ ($5$~yrs)  \ ($20$~yrs) & $10$~yrs \ ($5$~yrs)  \ ($20$~yrs) \\
  $q$      & Binary mass ratio & $0.1$ & $0.05$ & $1.0$ \\
\hline
\hline
  ${ \bf N_{\rm SBHB} }$     & {\bf Number of detectable ($\text{SNR}\geq2$) SBHBs }  & ${\bf 1.1}$ \quad ($0.3$) \quad ($3.1$) &  ${\bf 1.3}$ \quad ($0.4$) \quad ($3.8$) & ${\bf 0.8}$ \quad ($0.2$) \quad ($2.0$) \\
\end{tabular}
\caption{
Model parameters and the resulting number of \textit{Gaia}-detectable SBHBs (note that a 20 year mission lifetime requires a successor to \textit{Gaia}).}
\label{Table:params}
\end{table*}

\section{How many SBHBs could \textit{Gaia} detect?}

The angular scale of nearby sub-pc separation SBHBs is $\mathcal{O}(10)\mu$as.
The diffraction-limited imaging resolution of Gaia is $\sim10^4$ times larger.
While \textit{Gaia} cannot image sub-pc separation SBHBs, it does possess the
astrometric precision to detect $\sim10\mu$as centroid shifts in bright
sources.

We consider the case where only one SBH in the SBHB is luminous
\citep[\textit{e.g.}, Ref.][]{PG1302Nature:2015b}. Over the course of an
orbit, the position of the SBH, and thus the center of light, changes by a
characteristic value given by the semi-major axis of the
binary, $a$ (see \S \ref{S:Caveats} for further discussion). At 
angular-diameter distance $D_A(z)$, the orbital angular extent is
$\theta_{\orb} \approx a / D_A(z)$. \textit{Gaia} can detect
orbital motion if $\theta_{\orb}$ is greater than its
astrometric precision, and if the orbital period is shorter than twice the
mission lifetime.

\textit{Gaia}'s astrometric resolution can be parameterized by the brightness
and color of the source. Working in Johnson V-band magnitudes, we adopt an
average AGN $V-I_c=0.7$ based on the $r-i$ colors of nearby ($z\leq2.1$) SDSS
AGN \citep{PetersQcolor+2015}, and color correction equations
\citep{Jester+2005}, that yield a $V-I_c$ range of $0.3-1.1$. We use the fitting
formula from  Eqs.~(4-7) of Ref. \cite{GaiaII:2016} and the \textit{Gaia}
G-band to V-band conversion \citep{Evans_GtoV_DR2+2018} to compute the V-band
magnitude-dependent astrometric resolution of \textit{Gaia}. The astrometric
end-of-mission resolution, $\sigma_{\rm eom}$, is $9 \mu$as for a $m_V=13$
AGN \citep{GaiaDR2:astrmsoln:2018}.
This corresponds to a physical separation of $\sim 0.01$ pc at a distance of
$200$ Mpc, suggesting that \textit{Gaia} can probe sub-pc, GW-driven SBHBs if
they reside in nearby bright AGN.

Multiple works have considered exoplanet detection with \textit{Gaia}
\citep{Casertano+1995, BernsteinBastian:1995, Casertano+2008, Perryman+2014,
Ranalli+2018}. We draw on this body of work which shows that the relevant
quantity to consider for astrometric orbital detection is the signal-to-noise
ratio, $\text{SNR} = \theta_{\orb} / \sigma_{\rm sngle}$, where $\sigma_{\rm
sngle}$ is the precision for a single scan which we compute as $\sigma_{\rm
sngle} = \sqrt{70}/(2.15\times 1.2) \sigma_{\rm eom}$, the $5$-yr  end-of-
mission astrometric precision multiplied by the sky-position-averaged number
of scans per source (over $5$~yrs) and geometric sky averaging factors
\citep{Perryman+2014}. As shown in Ref. \cite{Ranalli+2018}, an SNR of $2.3$
($1.7$) is required to achieve a $\%50$ detection rate for a $5$-yr
($10$-yr) \textit{Gaia} mission, with a false-positive rate estimated at
$\mathbf{\lesssim 10^{-4}}$.  Hence in this work we adopt a minimum $\text{SNR}=2$
corresponding to a minimum detectable orbital angular size of
$\theta_{\min}=2\sigma_{\rm sngle}$. Next we compute the expected number of
such \textit{Gaia}-detectable SBHBs for both a $5$-yr mission and an extended
$10$-yr mission.

\subsection{Calculation}

We use the quasar luminosity function \citep[QLF;][]{HopkinsQLF+2007} to
derive the number of AGN per redshift $z$ and luminosity $L$. From $L$ and
$z$, and a bolometric correction to the V-band of $10$
\citep{RichardsQBCs:2006}, we find the corresponding V-band magnitude
$m_V(L,z)$, which gives the astrometric resolution, $\theta_{\min}$.
Combined with the redshift, this yields the minimum binary separation that
\textit{Gaia} can detect in that luminosity and redshift bin. At each
luminosity bin we derive a total binary mass from the assumption that the AGN
emits at a fraction of Eddington luminosity,
$L=f_{\Edd}L_{\Edd}(M)$. The minimum binary separation and the
binary mass yield the minimum binary orbital period for which \textit{Gaia}
could detect orbital motion,
\begin{equation}
P_{\min}(L,z) = \frac{2 \pi \left[\theta_{\min}(L,z) D_A(z)\right]^{3/2}}{\sqrt{G M(L,f_{\Edd})}}.
\label{Eq:Pmin}
\end{equation}
We adopt $f_{\Edd}=0.1$, motivated by an average value for bright AGN
\citep{KauffHeck:2009, SWM:2013_fEdds}.

We additionally require that the binary complete at least one orbit over
the course of the \textit{Gaia} mission. Otherwise orbital motion is
difficult to detect \citep{Ranalli+2018} or could be confused with
linear motion. The combined requirements constrain $P_{\min}(L,z)$ to be less
than a maximum time period $P_{\max}=10$~yrs ($5$~yrs) for a $10$-yr
($5$-yr) \textit{Gaia} mission.  We call AGN for which $P_{\min}(L,z)\leq P_{\max}$
`\textit{Gaia} targets'. This estimate, however, does not account for the
probability that an AGN harbors a SBHB at the desired orbital period. To
estimate this, we assume that a fraction $f_{\bin}$ of all AGN are triggered
by SBHBs. We then use the quasar lifetime $t_Q$ and the residence time of a
SBHB at orbital period $P$ to compute the fraction of $t_Q$ that a binary
spends at orbital periods below $P$ \citep[see, \textit{e.g.},][]{D'OrazioLoebVLBI:2018, HKM09}. The residence time due to GW emission
is,
\begin{eqnarray}     
t_{\rm{res}} &\equiv& \frac{a}{\dot{a}} = \frac{20}{256}  \left(\frac{P}{2 \pi}\right)^{8/3} \left(\frac{GM}{c^3}\right)^{-5/3} q^{-1}_s      
\label{Eq:FGWGas},
\end{eqnarray} 
for binary symmetric mass ratio $q_s\equiv 4q/(1+q)^2$, where
$q \equiv M_2/M_1\leq1$ and $M_1+M_2=M$. The probability for observing the
binary at orbital periods $\leq P$ is given by $\mathcal{F}(P, M, q_s) = {\rm
Min}\left[t_{\rm res}(P, M, q_s)/t_{\rm Q}, 1\right]$. We evaluate the
residence time at $P_{\min}$.

The total number of \textit{Gaia}-detectable SBHBs is,
\begin{eqnarray}
N_{\SBHB} &=& f_{\bin}\int^{\infty}_0 \left\{ 4 \pi \frac{d^2V}{dz d\Omega} \int^\infty_{\log{L_{\min}(z)}} \frac{d^2N}{d\log{L} dV} 
\mathcal{F}(P, M, q_s) \right. \nonumber \\
 &\times&  \mathcal{H}\left[ P_{\max} - P_{\min}(L,z) \right] \Bigg\}\ d\log{L} \ dz,
 \label{Eq:Ntot}
\end{eqnarray}
where $d^2N/d\log{L} dV$ is the pure-luminosity-evolution, double-power-law
QLF with redshift dependent slopes from Ref. \cite{HopkinsQLF+2007} (last row
of Table~3 labeled `Full'). $d^2V/dzd\Omega$ is the co-moving volume per
redshift and solid angle \citep{HoggCosmoDist:1999}, $\mathcal{H}$ denotes the
Heaviside function, $m_V(L_{\min},z)=21$, and we choose a fiducial quasar
lifetime $t_Q = 10^7$~yrs \citep{D'OrazioLoebVLBI:2018, PMartini:2004}.

\subsection{Results}
\label{S:Results}

\begin{figure*}
\begin{center}$
\begin{array}{c c c}
\hspace{-10pt}
\includegraphics[scale=0.37]{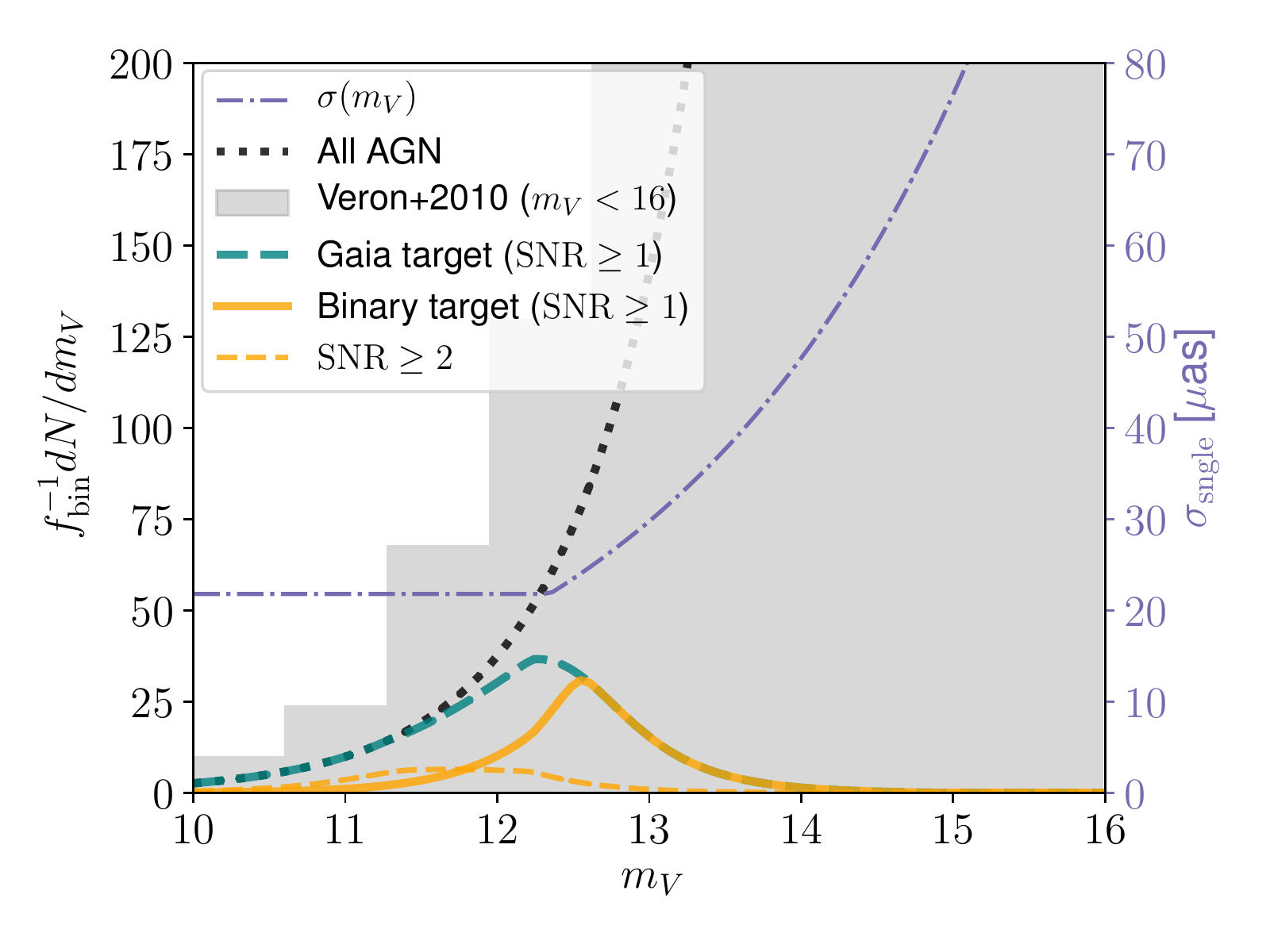} &
\includegraphics[scale=0.37]{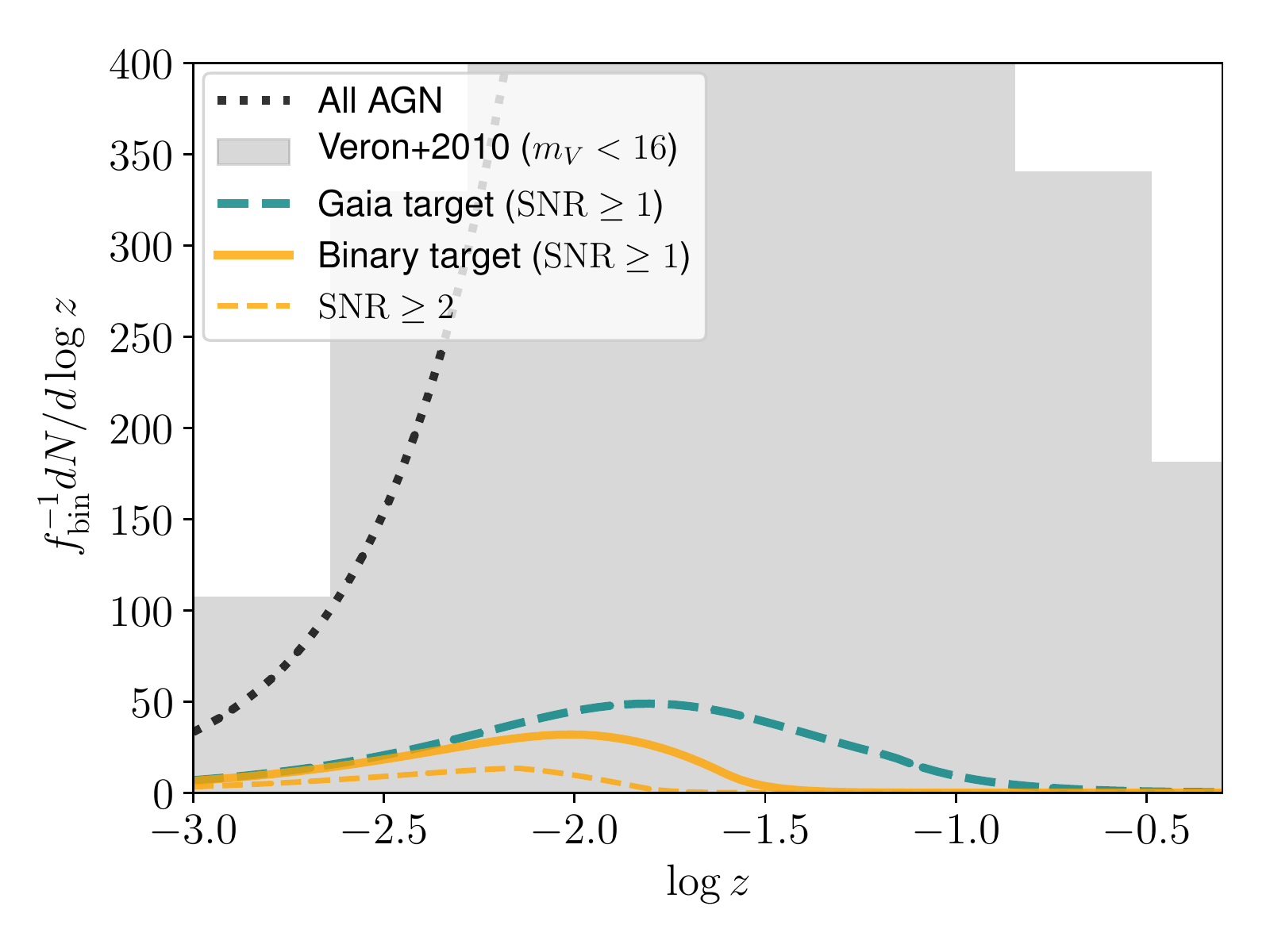} &
\includegraphics[scale=0.37]{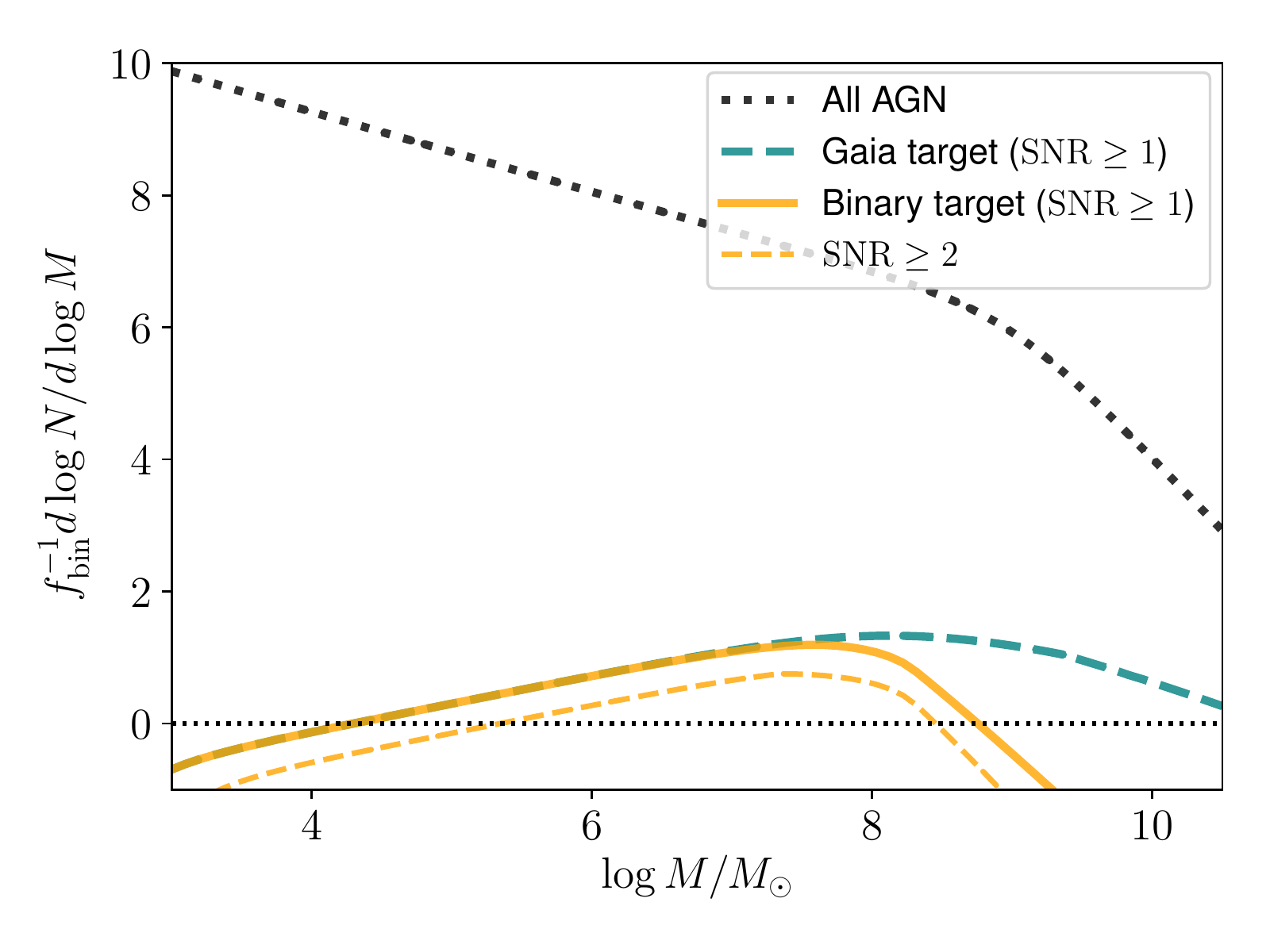}
\end{array}$
\end{center}
\vspace{-25pt}
\caption{
The number of AGN per V-band magnitude (left), log redshift (middle), and log
binary mass (right) for four different populations. The dashed-black line
shows all AGN. The teal-dashed line, labeled `\textit{Gaia}-target' shows only
the AGN for which the minimum  \textit{Gaia}-resolvable binary orbital period
(Eq.~\ref{Eq:Pmin}) is shorter than a $10$-yr \textit{Gaia} lifetime. The
orange lines weight the \textit{Gaia}-target distribution by the probability
for finding a SBHB at the required orbital period (with $f_{\bin}=1$ and
$SNR=1$ for visualization). The dashed-orange lines shows only those binaries
with a $\geq 50\%$ detection rate ($\text{SNR}\geq2$). The gray histograms count
known AGN with $m_V \leq 16.0$ \citep{VeronCat+2010}. In the left panel, the
purple dot-dashed line and corresponding right-vertical axis show the single-
scan astrometric precision of \textit{Gaia}.
}
\label{Fig:dNdx}
\end{figure*}

Table~\ref{Table:params} lists parameter choices and the resulting total
number of \textit{Gaia}-detectable SBHBs. For fiducial values, and a $10$-yr
\textit{Gaia} mission, $N_{\SBHB} \approx 11 f_{\bin}$. Thus, if the fraction
of SBHBs in local bright AGN is $f_{\bin} \gtrsim 0.1$, \textit{Gaia} has the
potential to find an SBHB during an extended, $10$-yr lifetime. Previous
studies have argued for a similar value of $f_{\bin}$ (typically $10\%$, which
is our fiducial value) based upon periodic variability searches in AGN
\citep{PG1302MNRAS:2015a, Charisi+2016}.

Table~\ref{Table:params} also lists our `optimistic' and `pessimistic'
parameter choices. In the optimistic case, $N_{\SBHB} \approx 13f_{\bin}$
SBHBs. In the pessimistic case, $N_{\SBHB} \approx 8f_{\bin}$ SBHBs. For each
case we also consider the benefit of $20$ years of observation (while
\textit{Gaia} cannot last that long, another 10 years of a successor mission
could fulfill this in the future \citep[\textit{e.g.}, Ref.][]{GaiaNIR:2016}).
Such an extended mission could result in up to $38 f_{\bin}$ putative SMBHB
detections.

Figure~\ref{Fig:dNdx} plots distributions of \textit{Gaia} SBHB candidates vs.
V-band magnitude, redshift, and binary mass. We show: (i) the total number of
AGN found from integrating the QLF (black-dotted line); (ii) the number of
`\textit{Gaia}-target' AGN, (teal-dashed line); (iii) `binary-targets',
including the probability $\mathcal{F}(M,P,q_s)$ for an AGN to contain a
binary at the desired orbital period (orange line); and (iv) the 
binary-targets with $\text{SNR} \geq 2$, for which $\geq 50\%$ of the
population will be detectable with a $\lesssim10^{-4}$ false-positive rate. The
teal and solid- orange lines (ii and iii) are drawn for $\text{SNR}\geq1$ in order to
more easily discuss the target population discussed below and to compare to
the $\text{SNR} \geq 2$ case. Integration under the dashed-orange lines and
multiplication by $f_{\bin}$ yields $N_{\SBHB}$ in Table \ref{Table:params}.
For reference, the gray histograms show the observed distribution of nearby
AGN with $m_V<16$ \citep{VeronCat+2010}. Note, however, that
magnitudes relevant for our study are those of the central point source,
presumably generated by accretion onto either component of a putative binary,
whereas magnitudes from \citet{VeronCat+2010} can include also the extended
host galaxy for such nearby systems. Hence the \citet{VeronCat+2010} catalog
may overestimate the number of nearby bright systems in the context of this
work.

The left panel of Figure~\ref{Fig:dNdx} displays the number of SBHBs per AGN
V-band magnitude. Comparing the teal-dashed line labeled `\textit{Gaia}
target' and the  black-dotted line (All AGN), we see that the orbital period
cut $P_{\min} \leq P_{\max}$ removes AGN with $m_V \gtrsim 12.5$. This is
because \textit{Gaia}'s resolution worsens for dimmer targets. To illustrate
this, the purple dot-dashed line plotted on the right vertical axis of the
left panel shows \textit{Gaia}'s single-scan astrometric precision
vs. $m_V$.

Comparison of the dashed-teal line with the solid-orange line shows that
brighter AGN in the `\textit{Gaia} target' distribution are less likely to
harbor a SBHB at the required orbital period $P_{\min}$. This is because
nearby, bright AGN correspond to more luminous AGN which correspond to AGN
with higher binary masses via the Eddington relation. At a fixed orbital
period, higher mass binaries inspiral more quickly and are hence less likely
to be found. Where the teal and orange curves overlap is where the binary
residence time is at least the quasar lifetime.

The dashed-orange line for $\text{SNR}\geq2$ binaries effectively represents a
population with a larger minimum orbital period. Hence there are fewer such
binaries that lie between this minimum and $P_{max}$. The dashed-orange line is
higher than the solid-orange line at bright magnitudes because the probability
$\mathcal{F}$ is larger due to a longer minimum orbital period. The 
dashed-orange line shows that for the fiducial case, the detectable SBHB distribution
peaks at $m_V = 12$, with an expectation value greater than $1 f_{\bin}$ for
AGN with $10.3 \leq m_V \leq 13$.

The middle panel of Figure~\ref{Fig:dNdx} displays the redshift distribution
of \textit{Gaia}-detectable SBHBs. The maximum-orbital-period cut removes
candidate AGN at all redshifts, while the binary-target distribution is
reduced in number from the \textit{Gaia}-target distribution at higher
redshifts. The latter is because SBHBs at higher redshift must be more
luminous in order for \textit{Gaia} to resolve orbital motion. Again, more
luminous AGN are associated with more massive SBHBs which merge more quickly.
The $\text{SNR}\geq2$ binaries (dashed-orange line) have a $\log{z}$
distribution peaking at $z\sim 0.01$ with expectation value $\geq1 f_{\bin}$
for $z\leq0.02$.

The right panel of Figure~\ref{Fig:dNdx} displays the distribution in binary
mass of \textit{Gaia}-detectable SBHBs. Comparison of the black-dotted and
teal-dashed lines shows that the highest fraction of AGN are removed from the
\textit{Gaia}-target distribution at lower binary masses. This is because
SBHBs with lower masses have much longer orbital periods for the same angular
separation and redshift. Again, the comparison of the solid-orange and teal-dashed
lines shows that the expectation value for the number of
\textit{Gaia}-detectable SBHBs also decreases for more massive binaries.
For fiducial parameter values, the $\text{SNR}\geq2$ binaries distribute
in $\log{M}$ with a peak at $M\sim3\times 10^7$ and expectation value $\geq1
f_{\bin}$ for $M\leq3 \times 10^8 \Msun$.

For optimistic (pessimistic) parameter values (Table~\ref{Table:params}), the
distributions peak at nearly the same magnitudes with a similar though
slightly increased (decreased) range, and extends to higher (lower) redshifts
$z\lesssim 0.02$ ($z\lesssim 0.01$), and higher (lower) binary masses $M
\lesssim 5 \times 10^8 \Msun$ ($M \lesssim 8 \times 10^7 \Msun$). For a
shorter, 5-year mission lifetime, the $\text{SNR}\geq2$ population peaks at a
slightly dimmer $m_V\sim11$ and has expectation value $\geq1 f_{\bin}$ for
$10.4 \leq m_V \leq12$, $z\leq 0.01$, and $M\leq4 \times 10^7 \Msun$.

\begin{figure}
\begin{center}
\includegraphics[scale=0.5]{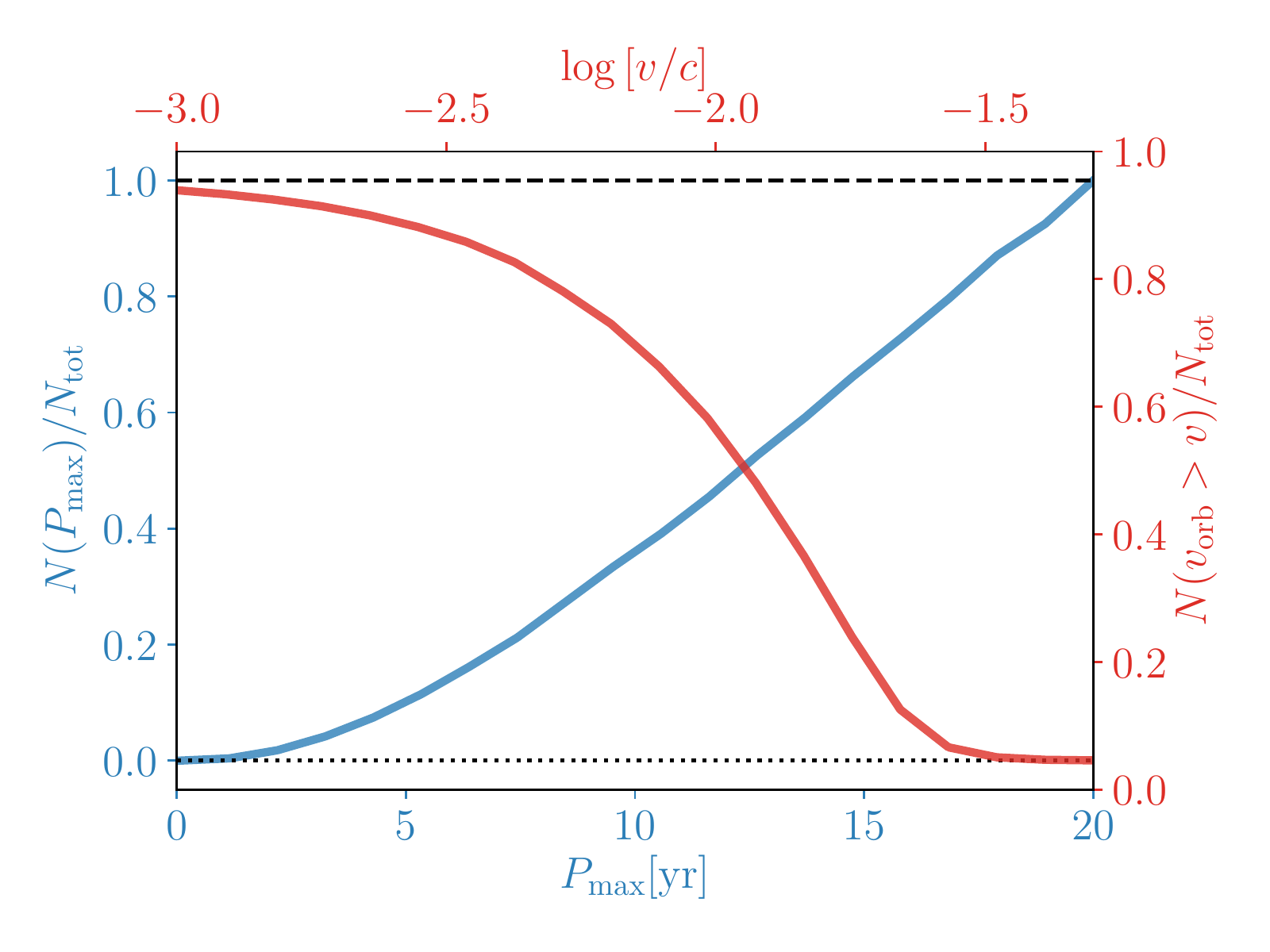}
\end{center}
\vspace{-25pt}
\caption{
\textit{Blue curve and left-bottom axes:} The fraction of \textit{Gaia}-detectable binaries vs. maximum detectable orbital period $P_{\max}$ (for a total number referenced to $P_{\max}=20$~yrs). \textit{Red curve and right-top axes:} The fraction of \textit{Gaia}-detectable binaries with orbital velocity of the secondary (mass ratio of $0.1$) greater than the labeled x-axis value. The orbital velocity in units of the speed of light approximates the amplitude of modulations induced by the orbital Doppler boost.
}
\label{Fig:PV}
\end{figure}

Cumulative distributions of $\text{SNR}\geq2$ binary targets in orbital period
and orbital velocity are plotted in Figure~\ref{Fig:PV}. The period
distribution (blue) shows the fraction of \textit{Gaia}-detectable SBHBs as a
function of $P_{\max}$.  We note that, while we assume a constant $5$-yr
mission-end resolution, this may change over the course of a longer mission
due to better PSF fitting, abut also due to possible instrument degradation.
The linear dependence of the period distribution indicates that the period
restriction $P_{\min} < P_{\max}$ dominates over the steeper $t_{\res}\propto
P^{8/3}$ residence-time dependence.

The velocity distribution (red) shows the number of \textit{Gaia}-detectable
SBHBs with orbital velocity $v_{\rm orb}/c$ above velocity $v/c$. This
quantity sets the fractional amplitude of photometric modulations caused by
the relativistic Doppler boost, given by $\Delta F_{\nu}/F_{\nu} \approx
(3-\alpha_{\nu}) v_{\orb}/c \cos{I}$, for specific flux $F_{\nu}$, $v_{\orb}/c
\ll 1$, inclination of the orbital plane to the line of sight $I$, and
frequency-dependent spectral slope $\alpha_{\nu}$ \citep[with typical values
$-2\lesssim \alpha_{\nu} \lesssim 2$; see Refs.][]{PG1302Nature:2015b,
Charisi+2018}. We compute $v_{\rm orb}/c$ as that of the secondary with
$q=0.1$.

Figure~\ref{Fig:PV} shows that \textit{Gaia}-detectable SBHBs will have
$v_{\rm orb}/c \lesssim 0.03$. Hence, for $\alpha_{\nu}=-2$, Doppler-induced
modulations will have $\Delta F_{\nu}/F_{\nu} \leq 5\%$, translating to
$\Delta m_V \leq 0.05$~mag amplitude modulations. \textit{Gaia}'s photometric
precision is better than $0.01$~mag at $m_V\lesssim14$ \citep{GaiaPhotom:2010,
GaiaDR2:2018} and could identify Doppler modulation coincident with
astrometric shifts of AGN optical regions. However, at $\sim$year timescales,
intrinsic AGN variability has often a higher amplitude than the maximum
$\Delta m_V = 0.05$~mag Doppler signal predicted here \citep{Macleod+2012},
and finding this signal without a \textit{Gaia} detection would be difficult.
If \textit{Gaia} identifies a SBHB candidate and its orbital period
astrometrically, then a targeted search for periodicity at the identified
orbital period, as well as further photometric monitoring beyond the lifespan
of \textit{Gaia}, could identify Doppler modulations, further validating the
SBHB interpretation.

\section{Discussion}

Binary motion can be uniquely identified and disentangled from linear motion.
Orbital motion in AGN would not be mistaken for a stellar binary because of
the much shorter orbital periods associated with more massive SBHs at the
measured orbital separation. Moreover, \textit{Gaia} measures high-resolution
spectra of objects with $V\leq15.5$ \citep{GaiaII:2016}, implying that AGN can
be identified unambiguously. Additionally, because \textit{Gaia} will observe
each bright object on the sky a median of $72$ times \citep[for the $5$-yr
mission;][]{GaiaII:2016}, candidate AGN spectra could be monitored for 
broad-line variations hypothesized to accompany SBHBs
\citep[\textit{e.g.},][]{NguyenBogdan+2018}, though \textit{Gaia}'s spectral
resolution may not be sufficient to detect such broad-line shifts and
variations. Broad-line monitoring from \textit{Gaia} or ground based
spectroscopic measurements along with multi-wavelength photometric monitoring
for binary-induced periodicity \citep[\textit{e.g.},][]{Graham+2015a,
Charisi+2016, LiuGezariPLCs+2016, PG1302MNRAS:2015a, PG1302Nature:2015b,
DZ:2017} could be used in tandem with \textit{Gaia} orbital tracking to prove
the existence of sub-pc separation SBHBs, and build a SBHB identification
ladder by studying the characteristics of confirmed SBHB-harboring AGN.

Because we predict the \textit{Gaia}-detectable SBHBs to lie in nearby, bright
AGN, future work should examine these known sources. Those exhibiting,
\textit{e.g.}, periodic variability should be given priority for examination
in the \textit{Gaia} dataset. If any \textit{Gaia} SBHB candidates are 
radio-loud, they can be targeted by mm-VLBI observatories that could simultaneously
track the orbital motion \citep{D'OrazioLoebVLBI:2018}, allowing orbital
tracking beyond the lifetime of \textit{Gaia} and offering insight into the
relation between radio and optical emission generated by SBHBs. Additionally,
SBHB orbital tracking can yield precise binary mass measurements, or even a
novel measurement of the Hubble constant \cite{D'OrazioLoebVLBI:2018}.

\subsection{Gravitational Waves}

The SBHBs detectable by \textit{Gaia} would be emitting GWs in the PTA
frequency band. As a consistency check, we follow Ref.
\cite{D'OrazioLoebVLBI:2018} and use the QLF to compute the corresponding
stochastic GW background (GWB). For simplicity and in difference from Ref.
\cite{D'OrazioLoebVLBI:2018}, we assume that the SBHBs are driven together
only by GW radiation and that $f_{\Edd}=0.1$. The resulting GWB falls a factor
of a few below the current PTA limits, consistent with previous studies
\citep[\textit{e.g.},][]{Kelley+2017}.

The most massive and nearby \textit{Gaia}-detectable SBHBs, have $M\sim
10^{8.5}\Msun$ and $z\sim0.01$ (Figure~\ref{Fig:dNdx}). Such an SBHB, with a
mass ratio of unity and an orbital period of less than $3$ years, could be
resolved as an individual source with a $\sim 13$ year PTA observation.
Determination of the orbital parameters and location on the sky by
\textit{Gaia} could aid PTA detection.

\subsection{Caveats}
\label{S:Caveats}

Throughout we have assumed that only one SBH is bright and that the light
centroid of the system moves a characteristic distance given by the orbital
semi-major axis. Depending on the relative masses and luminosities
of the two SBHs, however, this distance can vary. The motion of the light
centroid can be discerned from the difference between the fixed center of mass
of the binary and the center of light. Defining the Eddington-fraction ratio 
of the SBHs as $\xi \equiv f_{\Edd, 1}/f_{\Edd,
2}\leq 1$, we find that the change in light centroid over an orbit is,
\begin{equation}
\theta_{\orb} = \frac{2 a}{D_A(z)} \left( \frac{1}{1+q}  -  \frac{\xi}{1+\xi q} \right),
\end{equation}
simplifying to our fiducial value, $a/D_A(z)$, when only one SBH in an
equal mass binary is bright ($\xi=0$ and $q=1$). Orbital motion is
undetectable when both SBHs are accreting at the same fraction of Eddington.
However, this is a finely tuned case disfavored by previous work
\citep{Farris:2014, PG1302Nature:2015b}. If $\xi\leq1/3$, then our
adopted $\theta_{\orb}$ is reduced by less than a factor of two.

Because the primary sources for SBHB identification  with \textit{Gaia} are
nearby AGN, extended emission from a resolved nucleus could contribute to the
optical centroid. The extent of this complication must be studied further,
ideally for specific AGN candidates.

Another source of uncertainty lies in the assumption that an unknown fraction
$f_{\bin}$ of AGN are triggered by SBHBs. Additionally, our calculation relies
on the unknown rate at which SBHBs are driven to merger. We have only included
orbital decay due to GW radiation, as this is a process that must occur. But
gas accretion must also occur for the SBHs to be optically bright. To test the
affect of gas accretion on our results, we included a prescription for  gas-
driven orbital decay from Ref. \cite{D'OrazioLoebVLBI:2018}. Gas-driven decay
does not affect our result when occurring at less than the Eddington rate.
Furthermore, the binaries could stall before they make it to the small
separations considered here, in that case our binary probability prescription
is invalid and neither \textit{Gaia} nor any other technique will find very compact
SBHBs. However, detection of a SBHB with \textit{Gaia} could rule out that
possibility.

Since a detection of SBHBs would be the first of its kind, one may ask if the SNR cut
that we adopt from \citep{Ranalli+2018}, originally intended for astrometric
planet detection, yields a higher false-positive rate than desired for such
a task. Considering that there are only $\sim10^3$ bright nearby AGN
for which this detection method could be employed (\textit{e.g.} Figure
\ref{Fig:dNdx}), and that the $50\%$ detection rate is computed using a detection
criterion that was shown by \citet{Ranalli+2018} to yield a $\lesssim 10^{-4}$
false-positive rate, we view this as an acceptable minimum criteria for
motivating the possibility of SBHB orbital tracking.

We finally note, that the number of SBHBs with orbital separation larger than
the end-of-mission precision and $P_{\max}\leq20$~yrs is large, $\approx 440
f_{\bin}$. Such systems would move by an orbital separation that is resolvable
by \textit{Gaia} over its lifetime, but not necessarily resolved by the
single-scan precision for each of \textit{Gaia}'s $\sim70$ observations. While
not offering a definitive detection of a SBHB, such anomalous astrometric
measurements of AGN light centroids should be flagged for further
investigation.

\section{Conclusion}
We have shown that a 10~yr \textit{Gaia} mission has the capability to
astrometrically track the orbital motion of $\mathcal{O}(1)$ SBHBs in
bright ($m_V\lesssim13$), nearby ($z\lesssim0.02$) AGN. The discovery of SBHB
orbital motion over the next few years of the \textit{Gaia} mission would open
a new field of SBHB demography, generating an enormous boon for our
understanding of the mutual growth of SBHs and galaxies, evidence towards
resolving the final-parsec problem, the prospect of sources of gravitational
waves for PTAs, and a new method for calibrating cosmological distances
\citep{D'OrazioLoebVLBI:2018}. There is a strong incentive to analyze
astrometric data of bright, nearby AGN from \textit{Gaia} DR2 and onwards for
signatures of SBHB orbital motion.

\acknowledgements 
\vspace{-10pt}
We thank the anonymous referees for their useful comments that improved the
quality of this work. Financial support was provided from NASA through
Einstein Postdoctoral Fellowship award number PF6-170151 (DJD) and through the
Black Hole Initiative which is funded by a grant from the John Templeton
Foundation.

\bibliography{refs}
\end{document}